\documentclass[%
 reprint,
superscriptaddress,
 amsmath,amssymb,
 aps,
prx,
]{revtex4-2}

\usepackage{xcolor}
\usepackage{graphicx}
\usepackage{dcolumn}
\usepackage{bm}

\begin{document}


\title{Simultaneous Enantiomer-Resolved Ramsey Spectroscopy {\color{black}Scheme} for Chiral Molecules}
\author{Itay Erez}
\author{Eliana Ruth Wallach}
\affiliation{Solid State Institute, Technion-Israel Institute of Technology, Haifa, 3200003, Israel}
\affiliation{Physics Department, Technion-Israel Institute of Technology, Haifa, 3200003, Israel}
\affiliation{Schulich Faculty of Chemistry, Technion-Israel Institute of Technology, Haifa, 3200003, Israel}
\author{Yuval Shagam}
\email{yush@technion.ac.il}
\thanks{Career Advancement Fellow}
\affiliation{Solid State Institute, Technion-Israel Institute of Technology, Haifa, 3200003, Israel}
\affiliation{Schulich Faculty of Chemistry, Technion-Israel Institute of Technology, Haifa, 3200003, Israel}

\begin{abstract}
We {\color{black}theoretically} introduce a scheme to perform Ramsey spectroscopy on a racemic mixture of chiral molecules, simultaneously extracting the transition frequencies of the left- and right-handed molecules, known as enantiomers. By taking the difference between the enantio-specific frequencies, we isolate the weak force parity violation (PV) shift, which is predicted to break the symmetry between enantiomers.  {\color{black} To perform the scheme, we design a pulse sequence that creates enantio-specific superpositions in a three-level system using the enantiomer-dependent sign of the electric-dipole moment components' triple product. A delayed second pulse sequence completes the Ramsey interrogation sequence, enabling readout of the phase evolution for each enantiomer's transition through a separate quantum state.} Our technique overcomes the need to alternate between enantio-pure samples to measure PV.  We describe the advantages of the proposed method for precision metrology.
\end{abstract}

\maketitle

\begin{figure*}[t]
\includegraphics[width = 17.5cm]{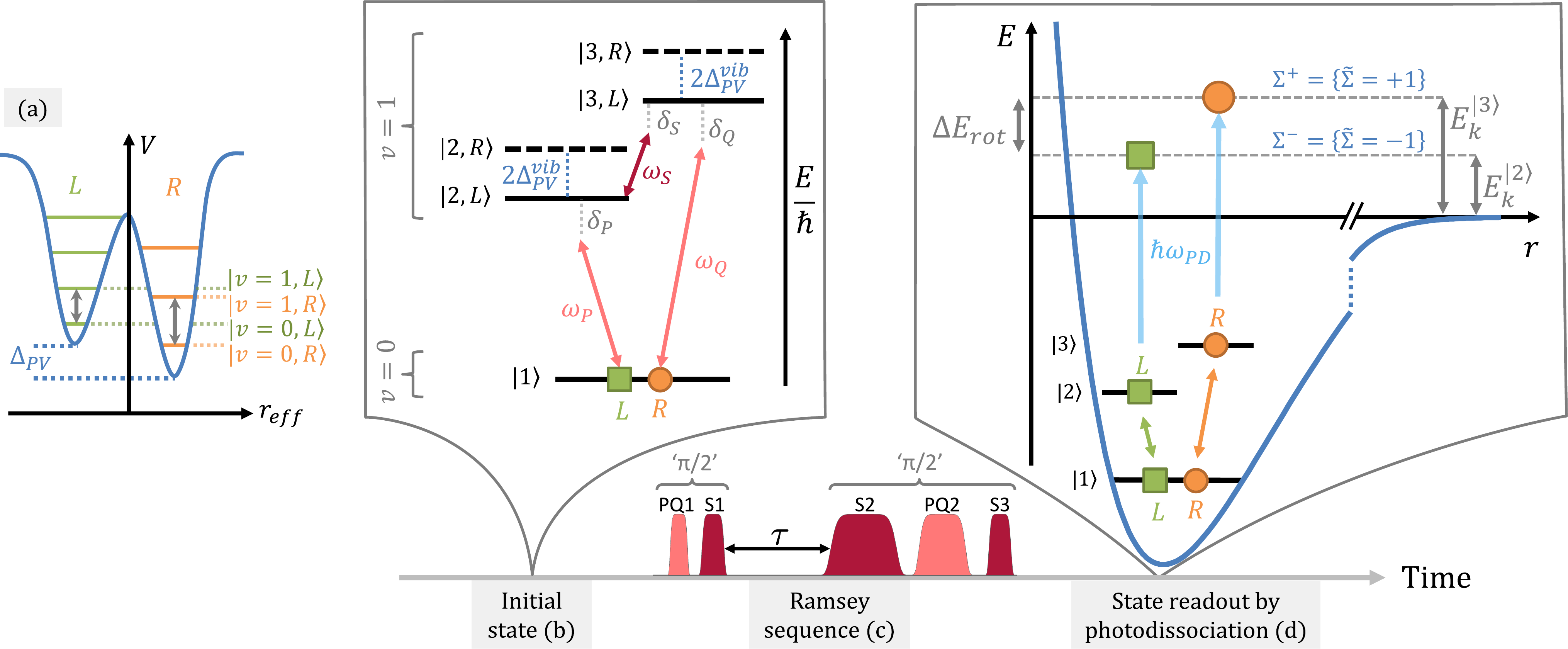}
\caption{Schematic outline of the Ramsey sequence for the case $\tilde{\Phi}=-1$. The chiral molecule is described by an asymmetric double well potential when including PV from weak interaction (a). The $R$ well is arbitrarily chosen to be deeper than the $L$ well. When considering the ground vibrational level $|1\rangle$ and two rotational states $|2\rangle$ and $|3\rangle$ in the first excited vibrational level, the slight difference in  frequencies of the two wells leads to an offset between the transitions of $L$ and $R$ molecules, {\color{black}denoted $2\Delta_{PV}^{vib}$} (b). {\color{black} The transitions between the three states are labeled P (pink), Q (pink) and S (dark red) and are driven by fields with the frequencies $\omega_P,\omega_Q,\omega_S$ that  are detuned by $\delta_P,\delta_Q,\delta_S$ from resonance, respectively (b). The fields are mutually orthogonally polarized.}  The molecules initially populate $|1\rangle$ and after the Ramsey sequence, {\color{black} which is composed of multiple individual pulses with varying application durations (as plotted in (c), see Appendix \ref{PulseDesc})},  part of the population of $L$ is transferred to $|2\rangle$ and part of $R$ to $|3\rangle$. {\color{black} The populations in $|2\rangle$ and $|3\rangle$ oscillate depending on the free evolution time, $\tau$, according to each enantiomer's detuning}. Readout of states $|2\rangle$ and $|3\rangle$ is possible by photo-dissociation, separating the two states by the kinetic energy of the photo-fragments (d). 
}
\label{fig:scheme}
\end{figure*}

The first observation of parity violation (PV)  was reported nearly 70 years ago  in the decay of $^{60}$Co \cite{Wu_1957}, shortly after parity conservation by the weak interaction was questioned \cite{LeeYang_1956}. The weak force  also has non-radioactive components known as neutral currents, which violate parity symmetry as detected in atomic transitions \cite{Wood1997,wieman2019atomic}. So far, PV has not been observed in chiral molecules, which are the prototypical example of mirror-image symmetry in nature.  Hund  originally modelled chiral molecules as a symmetric double-well potential \cite{Hund1927}, but the addition of PV creates an asymmetric double-well as shown in Fig.~\ref{fig:scheme}(a) \cite{Quack2012}.  The resulting vibrational transition frequencies of each enantiomer are shifted by $\pm\Delta_{PV}^{vib}$ \cite{Quack2000} [Fig.~\ref{fig:scheme}(b)]. Observation of PV in chiral molecules would confirm the prediction that the weak force is responsible for eliminating the so-called inter-enantiomer tunneling \cite{Quack2012}. It has even been hypothesized that this symmetry breaking seeds  the chiral excess exhibited in the chemistry of life, although this remains controversial \cite{YAMAGATA1966495,Zare2000,Quack2002,Quack2012,Fujiki2019,Senami2019}. Measurement of PV also has prospects beyond the Standard Model \cite{Safronova2018} such as in dark matter searches \cite{Berger2020}. 

The challenge of measuring  $\Delta_{PV}^{vib}$ lies in its tiny magnitude, where estimations for organic chiral molecules such as  {CHBrClF} are smaller than 10~mHz {\color{black} or a $10^{-16}$ shift relative to the vibrational transition} \cite{Quack2000} . However, calculations predict that $\Delta_{PV}^{vib}$ can be enhanced  by two orders of magnitude  for electronically excited states, due to the reduced cancellation of PV contributions from different orbitals \cite{Kuroda2022}. An alternative approach to increase the effect utilizes substitution of heavy metal nuclei in the molecule  \cite{Shwerdtfeger2003,Schwerdtfeger2004,Figgen2009,Figgen2010,Wormit2014}. In N$\equiv$UHFI for example,  $\Delta_{PV}^{vib}$ is calculated to rise to the order of 10~Hz {\color{black} or a $10^{-13}$ relative shift} \cite{Wormit2014}.

Multiple techniques have been used and proposed to measure PV in chiral molecules including {\color{black} using parity selection rules \cite{QUACK1986},} Fourier transform infrared (FTIR) spectroscopy \cite{Bauder1997,Daussy1999,Stoeffler2011}, microwave spectroscopy \cite{Satterthwaite2022}, vibrational Ramsey spectroscopy \cite{Cournol_2019}, NMR spectroscopy \cite{Eills2017} and matter-wave interferometry \cite{Stickler2021}. In {\color{black} most of} these techniques, the ability to synthesize enantiomer-enriched samples has been described as a key requirement \cite{Darquie2010,Schnell2011,Satterthwaite2022}. However, realizing this requirement is difficult for the molecules that are particularly appealing for PV measurement. Here we show how to avoid this requirement.

Many methods exist for enantiomer enrichment  with varying degrees of success  \cite{Domingos2018}{\color{black},  which can be employed to probe PV}. One scheme, known as three-wave mixing {\color{black} (3WM) \cite{Hirota2012}}, has particularly excellent prospects for efficient enrichment \cite{Patterson2013_2,Eibenberger2017,Perez2017,perez2018,Leibscher2019,Vitanov2019,Lee2022,Leibscher2022}. Its principles form the foundation of our work.  The scheme has been demonstrated to excite a sample of left ($L$) and right ($R$) handed molecules from an initial state $|1\rangle$ to either state $|2\rangle$ or $|3\rangle$ using the enantiomer-specific sign of the {\color{black} triple product of the} transition dipole moment components. 

Here we propose a fully differential scheme that leverages  racemic enantiomer mixtures to directly extract the PV signature. Instead of comparing transitions in separate enantio-pure samples, we {\color{black} theoretically design an identical procedure for both enantiomers, which combines Ramsey spectroscopy  and 3WM, } to simultaneously measure two transition frequencies in a racemic sample. This trait implies  support for common-mode noise rejection for the difference between the transitions, with the statistical sensitivity likely approaching the Standard Quantum Limit (SQL) as demonstrated in ref.\ \cite{Zhou2020}. This makes our scheme particularly appealing for precision measurement. Finally, we introduce a framework of experimental switches that isolate the PV contribution. {\color{black} We discuss the benefits of our procedure, which is embedded within the 3WM, 3-state framework, over the stringing of standard, 2-state, vibrational Ramsey spectroscopy followed by a microwave 3WM step.}

For our scheme, we developed a pulse sequence to perform  two Ramsey spectroscopies in a three-level system (Fig.~\ref{fig:scheme}). Each enantiomer is excited to a unique superposition using the sign difference between the scalar triple products of  their transition dipole moment components:  $\vec{\mu}_x^R\cdot\left(\vec{\mu}_y^R \times \vec{\mu}^R_z\right)=-\vec{\mu}_x^L\cdot\left(\vec{\mu}_y^L \times \vec{\mu}^L_z\right)$. Analogously to two-state Ramsey, a second excitation allows us to measure the energy difference of each superposition in separate final states. We apply the method to vibrational transitions  where PV is enhanced throughout the rest of this work. However, the method could be applied to ongoing  experiments with microwave rotational transitions for precision spectroscopy of PV \cite{Lee2022,Satterthwaite2022}.

For our procedure, three states are coupled by fields along the P, Q and S transitions [Fig.~\ref{fig:scheme}(b)] with mutually orthogonal polarizations to interact with all 3 {\color{black}electric} dipole components. We set the phase relation of the fields to
\begin{equation}
    \Delta\phi=\phi_P-\phi_Q+\phi_S=\pm\pi/2
    \label{eq:fieldphase}
\end{equation}
which is required to break the symmetry between states $|2\rangle$ and $|3\rangle$ in order to achieve perfect enantiomer separation \cite{Leibscher2019}.
The applied time-dependent interactions are $-\langle \vec{\mu}\cdot \vec{E}\rangle=2\hbar\Omega_X\cos(\omega_Xt+\phi_X)$, where $X~\in~\{P, Q, S\}$ and $\Omega_X$, $\omega_X$, $\phi_X$, and $\delta_X$  denote the Rabi rate, frequency, phase, and detuning from the resonant transition of each field.  {\color{black}The energy level diagram is} depicted  in Fig.~\ref{fig:scheme}(b). Applying the rotating wave approximation (RWA), the Hamiltonian describing the interaction for multiple configurations is 
\begin{equation}
    \label{eq:HRWA}
    \frac{H_\mathrm{RWA}}{\hbar} = 
    \begin{pmatrix}
        0 & \Omega_P & \Omega_Q\\
        \Omega_P & \delta -\delta_S/2 +\tilde{\kappa}\Delta_{PV}^{vib} & \tilde{\Phi}\tilde{\kappa}e^{i \pi/2}\Omega_S\\
        \Omega_Q & \tilde{\Phi}\tilde{\kappa}e^{-i \pi/2}\Omega_S &  \delta +\delta_S/2 +\tilde{\kappa}\Delta_{PV}^{vib}
\end{pmatrix}
\end{equation}
The Rabi rates $\Omega_P$, $\Omega_Q$ and $\Omega_S$ are time dependent and pulsed on individually [Fig.~\ref{fig:scheme}(c)]. We choose the detunings such that $\delta_P-\delta_Q+\delta_S=0$ and use the convenient notation of the average vibrational detuning $\delta = (\delta_P+\delta_Q)/2$. The handedness of the molecule is denoted by $\tilde{\kappa}$, where $\tilde{\kappa}(L)=-1$ and $\tilde{\kappa}(R)=+1$ such that the PV energy shift of an enantiomer is $\tilde{\kappa}\Delta_{PV}^{vib}$. To model the sign change of the transition dipole moment, $\tilde{\kappa}$ is applied without loss of generality to the $S$ transition $\tilde{\kappa}\langle 2| \vec{\mu}\cdot \vec{E}|3\rangle$ \cite{Leibscher2019}. The sign of the relative phase [Eq.~(\ref{eq:fieldphase})] is denoted by $\tilde{\Phi}=\pm1$ using the relation $e^{\pm i\pi/2}=\tilde{\Phi} e^{i\pi/2}$.

Our scheme begins with both $L$ and $R$ enantiomers populating state $|1\rangle$. Following Fig.~\ref{fig:ramsey}(a) for the ${\tilde{\Phi}=-1}$ ($\Delta \phi=-\pi/2$) case, the first `$\pi/2$' pulse sequence  creates the enantiomer-specific superposition states $|L\rangle=(|1\rangle-i|2\rangle)/\sqrt{2}$ and $|R\rangle=(|1\rangle-i|3\rangle)/\sqrt{2}$. {\color{black}This sequence is composed of two pulses: a simultaneous application of the fields $\Omega_P$ and $\Omega_Q$ (PQ1) followed by an application of the $\Omega_S$ field (S1).} After the molecules freely evolve for a duration $\tau$, we apply the second `$\pi/2$' pulse sequence to stop the phase evolution. {\color{black}This second sequence consists of an $\Omega_S$ pulse (S2), followed by $\Omega_P$ and $\Omega_Q$ applied together (PQ2), and finally another $\Omega_S$ pulse (S3) (see Appendix \ref{PulseDesc} for pulse areas and intermediate quantum states).} The resulting populations in states $|2\rangle$ and $|3\rangle$ ($\tilde{\Sigma}=-1, +1$ respectively) as a function of $\tau$ are depicted in Fig.~\ref{fig:ramsey}(b) and given by the general equations
\begin{equation}
N_{\tilde{\Phi},\tilde{\Sigma}} = \frac{\bar{N}}{2} \left[\cos\left(\delta_{\tilde{\Phi},\tilde{\Sigma}} {\tau}\right)+1\right]
\end{equation}
\begin{equation}
 \label{eq:energy}
 \delta_{\tilde{\Phi},\tilde{\Sigma}} = \delta+\tilde{\Sigma}\frac{\delta_S}{2}-\tilde{\Phi}\tilde{\Sigma}\Delta_{PV}^{vib}
\end{equation}
where $\bar{N}$ is the mean number of each enantiomer in the sample. We apply a realistic {\color{black}$\Delta_{PV}^{vib}= 2\pi\cdot10~\mathrm{Hz}$} as calculated for N$\equiv$UHFI \cite{Wormit2014} for example, which causes the slightly different frequencies for the populations of $|2\rangle$ and $|3\rangle$. These two frequencies mix in $|1\rangle$ where the $L$ and $R$ populations beat.

\begin{figure}
\includegraphics[width=8cm]{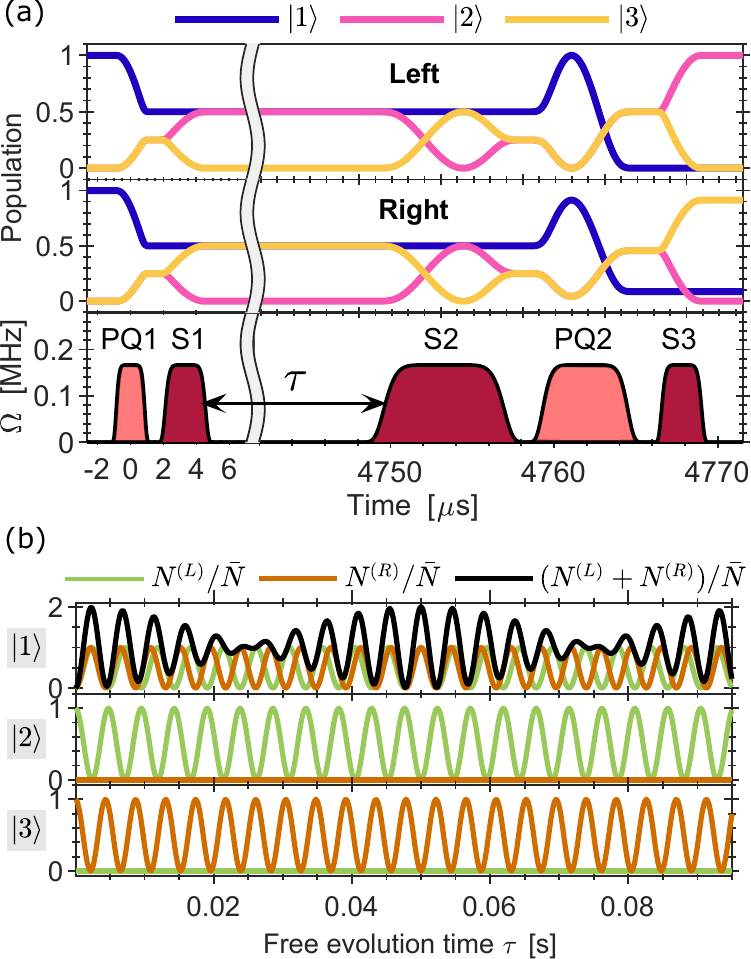}
\caption{The pulse sequence for a free evolution time of 4.7~ms (single fringe) is shown in (a) along with the time dependent magnitudes of the wavefunctions of states $|1\rangle$, $|2\rangle$ and $|3\rangle$ for $L$ and $R$ enantiomers. Here $\tilde{\Phi}=-1$, $\delta_S=0$ and $\delta \approx 1.4~\mathrm{kHz}$.  Identical simultaneous pulses along the P and Q transitions are denoted by the PQ symbol. The pulse areas of PQ1, S1, S2, PQ2 and S3 are $\pi/\sqrt{32}$, $\pi/4$, $3\pi/4$, $3\pi/\sqrt{32}$, and $\pi/4$ respectively {\color{black}(see Appendix \ref{PulseDesc})}.  While $L$ enantiomers undergo exactly one cycle, $R$ enantiomers are slightly past the top of the fringe due to the {\color{black}$\Delta_{PV}^{vib}=2\pi\cdot10$ Hz} difference. The resulting {\color{black} normalized} populations {\color{black} for each enantiomer, $N^{(L)}/\bar{N}$  and $N^{(R)}/\bar{N}$, in states}  $|1\rangle$, $|2\rangle$ and $|3\rangle$  of the Ramsey sequence  are plotted in (b) as a function of free evolution time $\tau$.}
\label{fig:ramsey}
\end{figure}

{\color{black} The populations in states $|2\rangle$ and $|3\rangle$ need to be measured for various values of $\tau$ in order to extract the detuning of each enantiomer's superposition ($\delta_{L}=\delta_{\tilde{\Phi}=-1,\tilde{\Sigma}=-1}$, $\delta_{R}=\delta_{\tilde{\Phi}=-1,\tilde{\Sigma}=1}$).}
{}
{\color{black} This can be achieved} by single-photon photo-dissociation of the molecule and separating the photo-fragments according to their kinetic energies, via velocity map imaging \cite{Eppink1997} for example [Fig.~\ref{fig:scheme}(d)].  If needed, the kinetic energy difference can be amplified by further excitation of the population in one of these states.

\begin{figure}
\includegraphics[width=8.6cm]{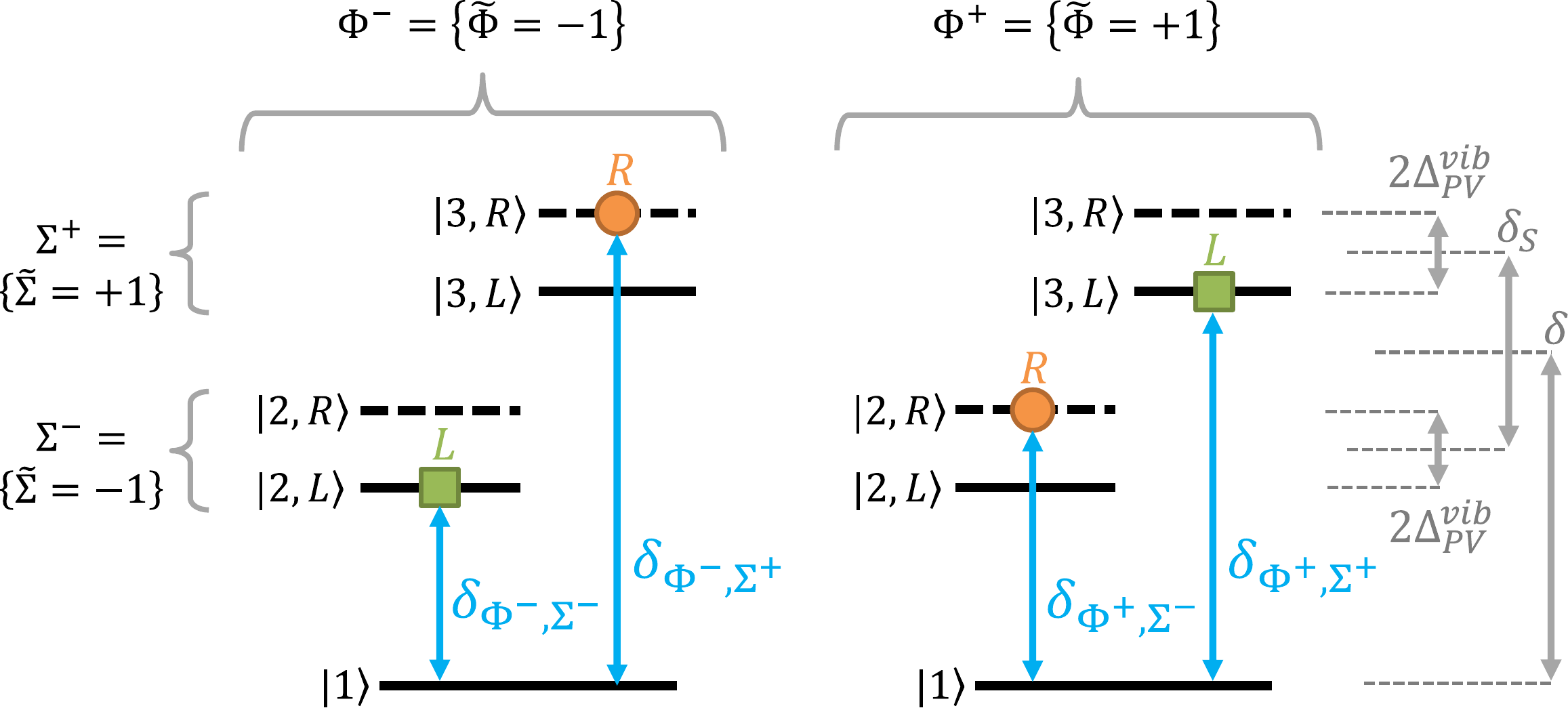}
\caption{The rotating frame level diagram is shown including the weak force PV component. The frequencies of the 4 transitions acquired during the experiment sequence are depicted by the blue arrows. Each arrow corresponds to a different experimental configuration of the `phase' and the detected `state' denoted by  $\tilde{\Phi}$ and $\tilde{\Sigma}$, respectively. The linear combinations of these switches form frequency channels [Eq.~(\ref{eq:lincomb})].}
\label{fig:channels}
\end{figure}

{\color{black} Since the difference between the oscillation frequencies of the populations of $|2\rangle$ and $|3\rangle$ is $\delta_S-2\tilde{\Phi}\Delta^{vib}_{PV}$ according to Eq.~(\ref{eq:energy}), an additional switch of the experimental configuration, ${\tilde{\Phi}}$, must be added to isolate $\Delta^{vib}_{PV}$.} When the relative phase of the fields is switched to ${\tilde{\Phi} = +1}$ ($\Delta \phi =\pi/2$), the roles of states $|2\rangle$ and $|3\rangle$ are reversed (Fig.~\ref{fig:channels}). In this case, the superposition states formed in the Ramsey sequence are $|L\rangle=(|1\rangle-i|3\rangle)/\sqrt{2}$ and $|R\rangle=(|1\rangle-i|2\rangle)/\sqrt{2}$. The frequencies read out by the Ramsey sequence for a specific relative phase $\tilde{\Phi}$ and state readout $\tilde{\Sigma}$ are given by Eq.~(\ref{eq:energy}), where $\tilde{\kappa} = -\tilde{\Phi}\cdot\tilde{\Sigma}$ encodes the chirality for a given configuration.

All four switch-states or configurations $\{\tilde{\Phi},\tilde{\Sigma}\}=\{\pm 1, \pm1\}$ must be measured to complete a data block (Fig.~\ref{fig:channels}). The linear combinations of these switch-states form frequency channels, which have intuitive physical meaning.
\begin{equation}
\label{eq:lincomb}2\pi
\begin{pmatrix}
      f_0 \\
      f_\Phi \\
      f_\Sigma \\
      f_{\Phi\Sigma}
\end{pmatrix}
\equiv
\frac{1}{4}
\begin{pmatrix}
+ & + & + & +\\
+ & + & - & -\\
+ & - & + & -\\
+ & - & - & +
\end{pmatrix}
\begin{pmatrix}
      \delta_{\Phi^+,\Sigma^+} \\
      \delta_{\Phi^+,\Sigma^-} \\
      \delta_{\Phi^-,\Sigma^+} \\
      \delta_{\Phi^-,\Sigma^-}
\end{pmatrix}
=
\begin{pmatrix}
      \delta \\
      0 \\
      \delta_S/2 \\
      -\Delta_{PV}^{vib} 
\end{pmatrix}
\end{equation}

To us, $f_{\Phi\Sigma}$ is the most important frequency channel  since it is both $\Phi$-odd and $\Sigma$-odd, thus isolating the PV component  of the two vibrational transitions of $L$ and $R$. The other channels are used to tune the experiment and to probe sources of systematic error. For example, the $f_{\Sigma}$ channel can be used to measure any $\Sigma$-odd effects between states $|2\rangle$ and $|3\rangle$ such as Zeeman shifts with different g-factors for the states {\color{black} (see Appendix \ref{magnetic})}. These can be corrected with a $\delta_S$ offset. 

When states $|2\rangle$ and $|3\rangle$ are measured simultaneously in a resolved manner, the two  $\Sigma$-odd channels, $f_\Sigma$ and $f_{\Phi\Sigma}$,  benefit from common mode noise suppression. This includes dissociation laser noise, which can be classified as proportional number noise, and magnetic field noise, which manifests as frequency noise. Number noise effects cancel to first order when taking the difference of ${N_{\tilde{\Phi},{\Sigma^+}}-N_{\tilde{\Phi},{\Sigma^-}}}$. Importantly, frequency noise effects are also suppressed for such a term when ${N_{\tilde{\Phi},{\Sigma^+}}}$ and ${N_{\tilde{\Phi},{\Sigma^-}}}$ are in-phase. Similar noise rejection has been demonstrated to achieve the SQL in a noisy photo-dissociation  measurement of the permanent electric dipole moment of the electron (eEDM) \cite{Zhou2020}. {\color{black} According to the SQL when perfect contrast is assumed, the uncertainty on $f_{\Phi\Sigma}(\propto\Delta^{vib}_{PV})$ scales as $(T\sqrt{N})^{-1}$, where $T$ is the coherence time and $N$ is the number of molecules detected. When considering a beam experiment \cite{Cournol_2019} or a trapped molecule experiment (as we plan), reasonable coherence times to expect are $T= 1$~ms or $T= 100$~ms, respectively. This means that detection of only 100M or 10k molecules, respectively,  would be needed to reach an uncertainty of 0.1~Hz as opposed to a much higher number depending on the technical noise if the SQL cannot be reached. This is far below the limit required to measure $\Delta^{vib}_{PV}$ for N$\equiv$UHFI for example. {\color{black} Drifts of the ambient conditions will be suppressed regardless of their rate in the proposed simultaneous measurement, as opposed to any implementation where the probed enantiomer is switched,  which would limit the bandwidth of the noise rejected, such as in ref.\ \cite{Satterthwaite2022} where switching enantiomers takes several minutes.}}

{\color{black} Alternatively, if simultaneous measurement is not possible,} the two states can also be read out sequentially by any state-selective technique such as fluorescence, state-selective photo-ionization/photo-dissociation, or state dependent chemistry. This slightly reduces the number of suppressed noise sources{\color{black}, but  significant noise reduction is maintained  in spite of the sequential readout since the Ramsey free evolution occurs simultaneously. Moreover, the simultaneous free evolution, relaxes the required relative stability to measure $\Delta_{PV}^{vib}$ to $10^{-9}$  for N$\equiv$UHFI as opposed to $10^{-13}$ when comparing independently probed vibrational transitions. This is because $\Delta_{PV}^{vib}$ is compared to $|\omega_P-\omega_S|$, the rotational splitting, and not the vibrational transition.}

In order to test our sensitivity to changes in the experiment parameters, we varied the relative areas of each of the pulses in the sequence. We define the contrast of the fringe in state $\tilde{\Sigma}$ as
\begin{equation}
 \label{eq:contrast}
 C_{\tilde{\Sigma}} = \left|N^{(L)}_{\tilde{\Sigma}}-N^{(R)}_{\tilde{\Sigma}}\right| / \bar{N}
\end{equation}
where $ N_{\tilde{\Sigma}}^{(\tilde{\kappa})}$ is the amplitude of the population oscillation of the enantiomer $\tilde{\kappa}$ in state $\tilde{\Sigma}$. We also define the leakage for the specific case of $\tilde{\Phi}=-1$  as 
\begin{equation}
 \label{eq:leakage}
\mathrm{Leakage}= \left(N_{|2\rangle}^{(R)}+N_{|3\rangle}^{(L)}\right)/2{\color{black}\bar{N}}
\end{equation}
which describes the average amplitude of the contaminant enantiomer that is not expected in each state for perfect pulses. 
For the case of $\tilde{\Phi}=1$ the leakage definition is $(N_{|2\rangle}^{(L)}+N_{|3\rangle}^{(R)})/2{\color{black}\bar{N}}$. 

\begin{figure}
\includegraphics[width=8.6cm]{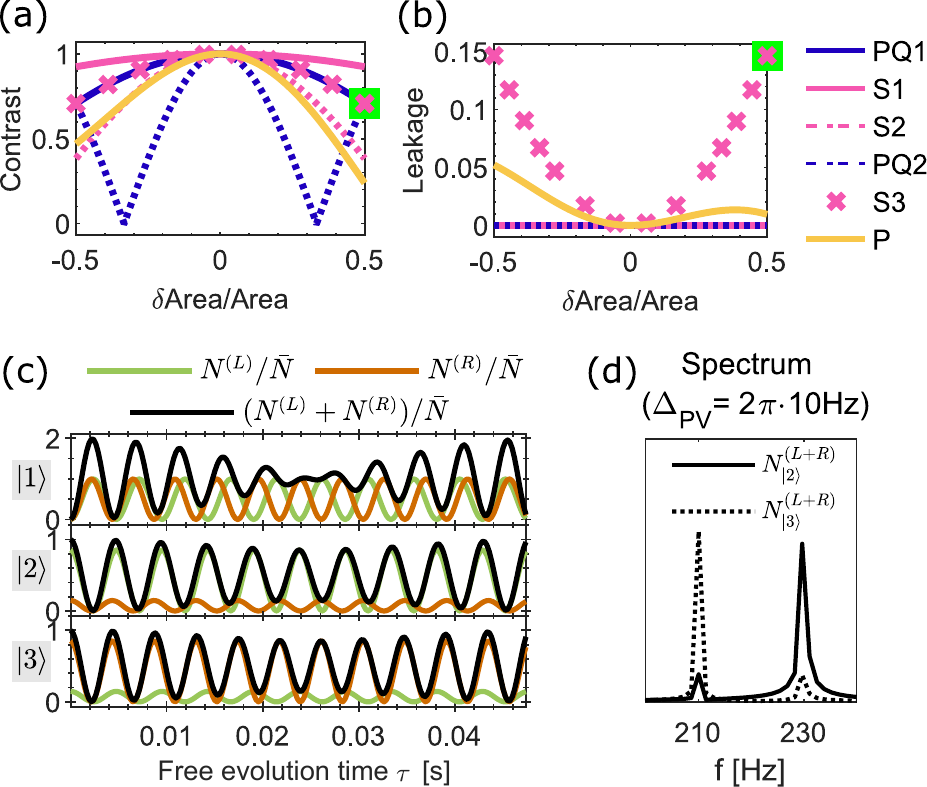}
\caption{Sensitivity of the contrast (a) and enantiomer leakage (b) due to an inaccurate area of each pulse. {\color{black} Additionally, P denotes the variation of the P pulse area alone with a constant Q pulse.} (c) A Ramsey sequence, showing significant oscillation contamination of the wrong enantiomer (leakage) for $\tilde{\Phi}=-1$, is created with a variation of 50\% in S3 [green squares in (a,b)]. This contamination can be observed in the spectra of the total populations  (d), but would only lead to an underestimation of $\Delta_{PV}^{vib}$. 
} 
\label{fig:sensitivity}
\end{figure}

{\color{black} While a low $C_{\tilde{\Sigma}}$ leads to a reduced precision of the measurement, a high leakage can lead to a systematic shift of $f_{\Phi\Sigma}(\propto\Delta^{vib}_{PV})$. However, we find that moderate stabilization of experiment parameters leads to minimal leakage and contrast loss.} The contrast and leakage are plotted in Fig.~\ref{fig:sensitivity}(a,~b) for a variation of each pulse area. When these variations are stabilized to within 10\%, the contrast is more than 0.88 and the leakage is less than 0.01. 

Fig.~\ref{fig:sensitivity}(c) shows a case with substantial leakage in the fringes due to a large 50\% error in the area of pulse S3. The spectrum observed in states $|2\rangle$ and $|3\rangle$ shows the opposite enantiomer frequency as a small feature on the sides of the main peaks [Fig.~\ref{fig:sensitivity}(d)]. This leakage may seem to be a detrimental systematic in the scheme, but would only lead to an underestimation of $\Delta_{PV}^{vib}$ with no phantom PV frequency appearing. A similar argument can be made if the $\tilde{\Sigma}$ state readout is not fully resolved. 

{\color{black} To further emphasize the advantages of our method, where the Ramsey scheme is embedded within the three-level system, we compare to an alternative-straightforward scheme where vibrational Ramsey spectroscopy is performed between two states (a ground and excited vibrational level) of the molecules in a racemic sample  followed by a subsequent microwave 3WM enantiomer-selective transfer step using 3 rotational states of the excited vibrational level.
Notably, both variants should have near perfectly enantiomer resolved frequencies since all the vibrationally excited states are unpopulated. While the two schemes have similar experimental complexity, both requiring at least one stable laser and at least one microwave field, our proposed embedded scheme has significant advantages.

The first important advantage to precision measurement stems from the benefits of monitoring possible sources of systematic uncertainty such as ambient magnetic fields. In both schemes, multiple experimental switches must be incorporated into linear combinations of the observed frequencies to isolate  $\Delta_{PV}^{vib}$ [Eq.\ (\ref{eq:lincomb})]. The other combinations are used to monitor other components of the Hamiltonian. In Appendix \ref{magnetic}, we show how an additional switch of the sign of $\delta_S$ can be used to measure an ambient magnetic field \textit{in situ}, due to the lever arm that arises from the different g-factors of the rotational states $|2\rangle$ and $|3\rangle$. This \textit{in situ} magnetometer type measurement is not possible for the straightforward non-embedded scheme since the same ground and excited states are used for both enantiomers. While auxiliary measurements with external magnetometers are possible to estimate the possible systematics, they are not as precise as those detected by the molecules themselves, which is crucial to precision measurement as discussed by Wu \textit{et al.}~\cite{Wu_2020}. Additional switches can be used to isolate other effects caused by the surroundings.

A second important advantage concerns maintaining the measurement precision on PV at the SQL or Quantum Projection Noise Limit  toward new physics searches \cite{Berger2020}, once a non-zero PV is hypothetically observed. Importantly, common-mode noise rejection only works if the two simultaneously measured sine waves are nearly in-phase \cite{Zhou2020}. If PV is large enough to cause a substantial phase shift between the enantiomers, then many noise sources can no longer be considered to be common-mode. While for a beam experiment such as Ref.~\cite{Cournol_2019}, the expected coherence time of $\sim$1~ms would induce a negligibly small phase shift, the shift would be large  when considering an experiment using trapped molecular ions, where the expected coherence time is $\sim$100~ms, even for PV shifts significantly smaller than $2\pi\cdot10$~Hz. However, in our scheme, purposefully applied $\delta_S$ corrections can be used to bring the two sine waves in-phase without waiting for a beating of the frequencies. This capability is crucial for searches for time variation in the signal such as in Dark Matter sensing \cite{Berger2020}, for example. There, the ability to bring the fringes in-phase at arbitrary, long evolution times enables the high frequency range to be measured at the SQL.

}

We now describe the experimental feasibility of our proposed method. The choice of the rotational levels $|2\rangle$, $|3\rangle$ in $v=1$ and level $|1\rangle$ in $v=0$ is not only taken to enhance the PV energy difference  relative to pure rotational transitions \cite{Cournol_2019,Quack2012}. It also avoids the thermal population contamination that exists in rotationally excited states, which harms the transfer contrast as pointed out by Koch and colleagues \cite{Leibscher2019} {\color{black} among others and  directly proven experimentally by Eibenberger-Arias and colleagues \cite{Lee2022}}.  These types of vibrational  transitions  are often situated in the mid to long-infrared wavelength range, where laser stabilization is challenging. Fortunately, adequately narrow lasers for coherent excitation in this range have already been demonstrated \cite{Cournol_2019}. Alternatively,  an  Optical Parametric Oscillator (OPO) laser can be used \cite{Vitanov2019}. Connecting pure rotational transitions such as states $|2\rangle$ and $|3\rangle$ is ubiquitous in {\color{black} microwave 3WM} schemes. {\color{black} Additionally, the scheme requires specific timings of pulses for the three transitions. In the microwave regime the timings can be set using microwave sources or amplifiers, but for the IR lasers, acousto-optic modulators, which are available in this wavelength range, (or another alternative) are required to implement the pulses. }

{\color{black} Our scheme assumes that the fields applied are sufficiently narrow to avoid decoherence for the target free evolution time $\tau$, and to set the phase condition of Eq.~(\ref{eq:fieldphase}). Interestingly, if the two IR transitions originate from the same laser via modulation at the microwave transition frequency by an electro-optic modulator, the phase  condition of Eq.~(\ref{eq:fieldphase}) can be achieved for a relatively broad linewidth laser by common-mode phase stability. This way, the separation of enantiomers would be guaranteed while the coherence time of the Ramsey sequence alone would be compromised by the linewidth, thereby reducing the susceptibility to shifts arising from leakage as discussed above.}

The scheme phases, wavelengths, and Rabi rates can be optimized experimentally despite the use of a racemic sample by using 2 sequential stages of {\color{black}3WM} with a total of 5 states, {\color{black} as was similarly demonstrated in P\'erez \textit{et al.}\ \cite{perez2018}}. The sign of  $\Delta_{PV}^{vib}$ can be measured using a small sample with enantiomeric excess. This is a less stringent requirement than using such a sample for the entire experiment. Alternatively,  concatenation of other methods to our scheme, such as Coulomb explosion \cite{pitzer2013}, can be used to determine the absolute configuration of the molecules \cite{Domingos2018}. To our knowledge, measuring $\Delta\phi$ for the fields and comparing the sign of the transition dipole moment components is not reliable in extracting the molecule's absolute configuration \cite{Domingos2018}.

We have presented a framework to perform PV precision measurement in a racemic sample of chiral molecules that is readily applicable to existing experiments using vibrational spectroscopy \cite{Cournol_2019} as well as pure rotational {\color{black}3WM} \cite{Satterthwaite2022,Lee2022}.  The differential scheme benefits from common-mode noise rejection such that statistical uncertainty estimations at the SQL are realistic. Our own plan is to realize this method with charged chiral molecules, which can be trapped, facilitating measurements at long coherence times \cite{Zhou2020}. Ionized versions of chiral molecules have also been suggested to substantially enhance PV \cite{Senami2019}. Our upcoming work discusses several chiral molecular ion candidates for PV measurement \cite{Landau2022}. 

\begin{acknowledgements} We thank M.\ Kr\"uger and Y.\ Soreq for careful reading of the manuscript {\color{black}and B. Darqui\'e for helpful discussions}. Y.S.\ is thankful for support from the Israel Science Foundation Grant No.\ 1142/21 and the Council for Higher Education Support Program for Hiring Outstanding Faculty Members in Quantum Science and Technology in Research Universities. E.R.W.\ acknowledges the support of the Technion Excellence Program for undergraduate students. 
\end{acknowledgements}

\appendix

{\color{black} \section{Pulse sequence description} \label{PulseDesc} To explain the pulse sequences used to create the composite `$\pi/2$' pulses, we describe a specific case of the Hamiltonian [Eq.\ (\ref{eq:HRWA})], where $\phi_P=\phi_Q$ such that $\Delta\phi=\phi_S = -\pi/2$ ($\Tilde{\Phi}=-1$) and $\delta_S =0$. This simplification isolates the $S$ pulses as those that discriminate between $L$ and $R$ molecules for a clearer description. Following Fig.\ \ref{fig:ramsey}(a), the first `$\pi/2$' pulse sequence begins with a simultaneous application of $\Omega_P$ and $\Omega_Q$ (denoted PQ1), each with an area of $\pi/\sqrt{32}$ to create the superposition 
\begin{equation*}
|L\rangle=|R\rangle=\frac{1}{\sqrt{2}}|1\rangle-\frac{i}{{2}}(|2\rangle+|3\rangle).
\end{equation*}
To complete the first `$\pi/2$' sequence an S pulse with an area of $\pi/4$ is applied (denoted S1), separating the enantiomers in $|2\rangle$, $|3\rangle$ to the unique superposition states $|L\rangle=(|1\rangle-i|2\rangle)/\sqrt{2}$ and $|R\rangle=(|1\rangle-i|3\rangle)/\sqrt{2}$. Both states evolve freely for a time $\tau$ to 
\begin{align*}
|L\rangle=(|1\rangle-i|2\rangle e^{-i\delta_L\tau})/\sqrt{2}\\
|R\rangle=(|1\rangle-i|3\rangle e^{-i\delta_R\tau})/\sqrt{2}
\end{align*}
before the second `$\pi/2$' pulse sequence is applied to translate the phase evolution of each enantiomer to oscillating populations in $|2\rangle$, $|3\rangle$ separately. The second `$\pi/2$' pulse sequence begins with an S pulse of area $3\pi/4$ (denoted S2) that reverses the effect of S1, by completing its area to $\pi$ and leading to the state: 
\begin{align*}
&|L(R)\rangle=\frac{1}{\sqrt{2}}|1\rangle+\frac{i}{2}\left(|2\rangle+|3\rangle\right) e^{-i\tau\delta_{L(R)}}
\end{align*}
Here the two wavefunctions are written in a shortened notation for $L$ with the corresponding $R$ wavefunction in parentheses. The S2 pulse symmetrizes the overall state relative to $|2\rangle$ and $|3\rangle$ before the second simultaneous PQ pulse (denoted PQ2) is applied. The PQ2 pulse has an area of $3\pi/\sqrt{32}$ and once applied the resulting state becomes:
\begin{align*}
|L(R)\rangle=\sin{\left(\tau\delta_{L(R)}/2\right)}|1\rangle+\frac{\cos{\left(\tau\delta_{L(R)}/2\right)}}{\sqrt{2}}\left(|2\rangle+|3\rangle\right)
\end{align*}
Alternatively, an identical pulse to PQ1 could be applied, but the resulting oscillations would be shifted by a  $\pi$ phase offset.
Finally, the sequence is completed with an S pulse (denoted S3) that is identical to S1 which transcribes the progression of the phase evolution to states $|2\rangle$ and $|3\rangle$ for $L$ and $R$ enantiomers respectively. The resulting states and populations are
\begin{align*}
&|L\rangle=\sin{\left(\tau\delta_{L}/2\right)}|1\rangle+\cos{\left(\tau\delta_{L}/2\right)}|2\rangle
\\
&|R\rangle=\sin{\left(\tau\delta_{R}/2\right)}|1\rangle+\cos{\left(\tau\delta_{R}/2\right)}|3\rangle
\\
&\implies\left\|\langle2|L\rangle\right\|^2=\frac{1+\cos\left(\tau\delta_{L}\right)}{2}=\frac{1+\cos\left(\tau(\delta-\Delta_{PV}^{vib})\right)}{2}
\\
&\implies\left\|\langle3|R\rangle\right\|^2=\frac{1+\cos\left(\tau\delta_{R}\right)}{2}=\frac{1+\cos\left(\tau(\delta+\Delta_{PV}^{vib})\right)}{2}
\end{align*}
Here we have substituted the individual enantiomer's detuning from its' respective state, $\delta_L$ and $\delta_R$, by the specific detuning of the switch state according to Eq. (\ref{eq:energy}).
Fig.~\ref{fig:ramsey}(a) shows a single fringe cycle for L molecules ($\tau = 2\pi /\delta_L$) with the resulting populations 
\begin{align*}
\left\|\langle2|L\rangle\right\|^2=1,~~ \left\|\langle3|R\rangle\right\|^2=\left(1+\cos\left(2\pi\frac{ 2\Delta_{PV}^{vib}}{\delta_L}\right)\right)/{2}
\end{align*}
This case shows how the PV signature can be isolated, but the scaling is quadratic. Naturally, for a precision measurement the side of the fringe would be preferred such that $\tau = \left(2\pi n+\frac{\pi}{2}\right)/\delta$ should be chosen, for an integer n. The resulting populations are
\begin{align*}
\left\|\langle2|L\rangle\right\|^2 \left(\left\|\langle3|R\rangle\right\|^2\right)=\left[1-\sin\left(\left(2\pi n + \frac{\pi}{2}\right)\frac{ \tilde{\kappa}\Delta_{PV}^{vib}}{\delta}\right)\right]/{2}
\end{align*}
where $\Tilde{\kappa}$ describes the handedness of the molecule according to $L$ and $R$. To also enjoy the benefits of noise cancellation $n$ should be chosen such that the two enantiomers' fringes are in-phase.

While we described the pulse pattern for a specific case, it also works for $\delta_S\ne 0$ and $\phi_P \ne \phi_Q$ with the $\tilde{\Phi}$ switch needed to isolate $\Delta_{PV}^{vib}$ as shown in Eq (\ref{eq:lincomb}).

 }

{\color{black}
\section{\textit{In situ} magnetic field measurement}
\label{magnetic}
An ambient magnetic field, $B_a$, can shift the energies of the three levels involved in our scheme. Such a shift is important to characterize in a precision measurement to estimate systematic uncertainty \cite{Wu_2020}. Here we show that the embedded simultaneous Ramsey scheme we propose enables an \textit{in situ} measurement of the  ambient magnetic field $B=B_a$, and facilitates its zeroing. The capability arises from the different sensitivities of states $|2\rangle$ and $|3\rangle$ to an external magnetic field. Different Land\'e g-factors and  Zeeman sub-levels, $m$, correspond to different magnetic moments $\mu_2=\mu_B g_2 m_2$ and $\mu_3=\mu_B g_3 m_3$ for $|2\rangle$ and $|3\rangle$ respectively. Here  $\mu_B$ is the Bohr magneton. We define $\mu_m = \frac{\mu_B}{2}(g_3 m_3 + g_2 m_2)$ and $\mu_\Delta = \frac{\mu_B}{2}( g_3 m_3 - g_2 m_2)$ as the mean and difference magnetic sensitivities, such that the magnetic moments of states $|2\rangle$ and $|3\rangle$ are $\mu_{\Tilde{\Sigma}}=\mu_m+\Tilde{\Sigma} \mu_\Delta$, using the $\Tilde{\Sigma}$ switch to distinguish between the two states. The shift in state $|1\rangle$ is not considered since it is common for the two excited states and is the initial state for both enantiomers. The Hamiltonian including the Zeeman shift  is
\begin{widetext}

\begin{equation}
    \label{eq:HRWA_B}
    \frac{H_\mathrm{RWA}^Z}{\hbar} = 
    \begin{pmatrix}
        0 & \Omega_P & \Omega_Q\\
        ~ & \delta -\delta_S/2 +(\mu_m-\mu_\Delta) B +\tilde{\kappa}\Delta_{PV}^{vib} &
        
        \tilde{\Phi}\tilde{\kappa}\Omega_S e^{i \pi/2}\\
        c.c &
        
        ~ & 
    
        \delta +\delta_S/2 +(\mu_m+\mu_\Delta)B +\tilde{\kappa}\Delta_{PV}^{vib}
\end{pmatrix}
 \end{equation}
 \end{widetext}

To isolate $B$, we add a switch of the sign of the microwave detuning $\delta_S$ such that $\delta_S\rightarrow\Tilde{S}\delta_S$ where $\Tilde{S}=\pm1$. To maintain the condition $\delta_P-\delta_Q+\delta_S=0$ the additional transformation of $\delta_P\rightarrow\Tilde{S}\delta_P$ and $\delta_Q\rightarrow\Tilde{S}\delta_Q$ must accompany the switch (note that $\delta\rightarrow\tilde{S}\delta$). The frequency of each switch state including the magnetic field is:
\begin{equation}
    \label{eq:energy_B}
    \delta_{\tilde{\Phi},\tilde{\Sigma},\tilde{S}} = \tilde{S}\delta+\tilde{\Sigma}\tilde{S}\frac{\delta_S}{2}+(\mu_m+\Tilde{\Sigma} \mu_\Delta)B -\tilde{\Phi}\tilde{\Sigma}\Delta_{PV}^{vib}
\end{equation}

For a magnetic field $B=0$  and the case $\Tilde{S}=1$, Eq.~(\ref{eq:energy}) is retrieved.  When the new switch $\Tilde{S}$ is added to the two previous switches $\tilde{\Sigma}$ and $\tilde{\Phi}$, the total number of frequency channels and linear combinations grows to $2^3$. These are given by:

\begin{widetext}
\begin{equation}
\label{eq:lincomb_B}
2\pi
\begin{pmatrix}
      f_0 \\
      f_\Phi \\
      f_\Sigma \\
      f_{\Phi\Sigma} \\
      f_S \\
      f_{\Phi S} \\
      f_{\Sigma S} \\
      f_{\Phi\Sigma S}
\end{pmatrix}
\equiv
\frac{1}{8}
\begin{pmatrix}
+ & + & + & + & + & + & + & +\\
+ & + & + & + & - & - & - & -\\
+ & + & - & - & + & + & - & -\\
+ & + & - & - & - & - & + & +\\
+ & - & + & - & + & - & + & -\\
+ & - & + & - & - & + & - & +\\
+ & - & - & + & + & - & - & +\\
+ & - & - & + & - & + & + & -\\
\end{pmatrix}
\begin{pmatrix}
      \delta_{\Phi^+,\Sigma^+,S^+} \\
      \delta_{\Phi^+,\Sigma^+,S^-} \\
      \delta_{\Phi^+,\Sigma^-,S^+} \\
      \delta_{\Phi^+,\Sigma^-,S^-} \\
      \delta_{\Phi^-,\Sigma^+,S^+} \\
      \delta_{\Phi^-,\Sigma^+,S^-} \\
      \delta_{\Phi^-,\Sigma^-,S^+} \\
      \delta_{\Phi^-,\Sigma^-,S^-} \\
\end{pmatrix}
=
\begin{pmatrix}
      \mu_m B  \\
      0 \\
      \mu_\Delta B \\
      -\Delta_{PV}^{vib} \\
      \delta \\
      0 \\
      \delta_S/2 \\
      0 \\
\end{pmatrix}
\end{equation}
 \end{widetext}

Importantly, the ambient magnetic field does not affect the $f_{\Phi\Sigma}$ channel, which rejects Zeeman shifts, even if they are state-dependent. Moreover, the magnetic field $B$ is measured in a resolved manner since it is the leading term in the $f_{\Sigma}$ channel, proportional to $\mu_\Delta$ as well as in the $f_{0}$ channel, which is proportional to $\mu_m$. This enables the measurement of the average vibrational detuning $\delta$ through the $f_S$ channel. Preferably, $f_{\Sigma}$ and $f_{0}$ channels can be used to apply a correction field of opposite sign $B_c=-B_a$ to zero the total magnetic field the molecules experience and prevent other systematic effects it may cause.

The procedure used to null the ambient magnetic fields can be employed to directly monitor and suppress other noise sources to isolate the PV signal by adding the relevant switch to the experiment. This procedure is possible, because the scheme is embedded within the 3-state manifold, since the two excited states used for the spectroscopy are likely to have different sensitivities to ambient effects. As discussed in the main text, if  Ramsey spectroscopy between a ground and excited vibrational state is applied to a racemic sample, most external effects will cause similar shifts to both enantiomers and preclude their monitoring. While such shifts will likely be suppressed to first order, this is not guaranteed for all effects and thus monitoring them is beneficial. 
}

\bibliography{apssamp}

\end{document}